\documentclass[twocolumn,aps,prl,showpacs,amsfonts]{revtex4-1}

\usepackage{amsfonts}
\usepackage{amsmath}
\usepackage{amssymb}

\begin{document}

\title{Quantum contextuality for rational vectors}

\author{Ad\'an Cabello}
 \email{adan@us.es}
 \affiliation{Departamento de F\'{\i}sica Aplicada II, Universidad de
 Sevilla, E-41012 Sevilla, Spain}

\author{Jan-{\AA}ke Larsson}
 \email{jan-ake.larsson@liu.se}
 \affiliation{Institutionen f\"or Systemteknik, Link\"opings
 Universitet, SE-581 83 Link\"oping, Sweden}

\date{\today}



\begin{abstract}
The Kochen-Specker theorem states that noncontextual hidden
variable models are inconsistent with the quantum predictions
for every yes-no question on a qutrit, corresponding to every
projector in three dimensions. It has been suggested [D. A.
Meyer, Phys. Rev. Lett. \textbf{83}, 3751 (1999)] that the
inconsistency would disappear when restricting to projectors on
unit vectors with rational components; that noncontextual
hidden variables could reproduce the quantum predictions for
rational vectors. Here we show that a qutrit state with
rational components violates an inequality valid for
noncontextual hidden-variable models [A. A. Klyachko {\em et
al.}, Phys. Rev. Lett. \textbf{101}, 020403 (2008)] using
rational projectors. This shows that the inconsistency remains
even when using only rational vectors.
\end{abstract}



\maketitle


The Kochen-Specker theorem from 1967 \cite{KS67} states that
the quantum predictions from a three-dimensional quantum system
(a qutrit) are inconsistent with noncontextual hidden
variables. The proof uses 117 directions in three dimensions,
arranged in a pattern such that they cannot be colored in a
particular manner, see \cite{KS67} for details. Later proofs
use less directions, but one common feature (in the
three-dimensional versions) is that the set of unit vectors
includes irrational components. It was noted in \cite{Meyer99}
that the Kochen-Specker proof needs these irrational vectors to
be completed. Indeed, when using only the rational subset of
vectors, the set is colorable in the manner required by quantum
mechanics. It was also suggested \cite{Meyer99} that, for this
reason, the inconsistency between quantum mechanics and
noncontextual hidden variables disappears, and that quantum
mechanics can be imitated by noncontextual hidden variable
models restricted to rational vectors.

It has been recently shown \cite{KCBS08} that the following
inequality is a necessary and sufficient condition for qutrit
noncontextual hidden variables, for measurements $A_i$ with
possible outcomes $-1$ and $+1$, such that $A_i$ and $A_{i+1}$
(modulo 5) are compatible:
\begin{equation}
\sum_{i=0}^4 \langle A_i A_{i+1} \rangle \ge -3.
 \label{KCBS}
\end{equation}
Using the rational qutrit state
\begin{equation}
\langle\psi| = \left(\frac{354}{527},\frac{357}{527},-\frac{158}{527}\right),
\end{equation}
and the observables
\begin{equation}
A_i = 2 |v_i\rangle \langle v_i| - \openone,
\end{equation}
associated to the rational vectors
\begin{subequations}
\label{rationalvectors}
\begin{align}
\langle v_0| = & \left(1,0,0\right),\\
\langle v_1| = & \left(0,1,0\right),\\
\langle v_2| = & \left(\frac{48}{73},0,-\frac{55}{73}\right),\\
\langle v_3| = & \left(\frac{1925}{3277},\frac{2052}{3277},\frac{1680}{3277}\right),\\
\langle v_4| = & \left(0,\frac{140}{221},-\frac{171}{221}\right),
\end{align}
\end{subequations}
we obtain a value of $-3.941$ for the left-hand side of
\eqref{KCBS}, which deviates very little from the maximum
violation at $-3.944$. Thus, even when using only rational
vectors, the inconsistency is not nullified. The violation
shows that the (physical content of) the Kochen-Specker theorem
remains, namely, that the quantum-mechanical predictions cannot
be reproduced by noncontextual hidden variables.


\end{document}